\documentclass[aps,twocolumn,showpacs,letterpaper,footinbib]{revtex4}
\usepackage[dvips]{graphicx}
\usepackage{amsmath}
\usepackage{amssymb}
\usepackage{dcolumn}
\usepackage{pifont}

\begin{document}
\title{Fluidization of a vertically oscillated shallow granular layer}

\author{Jennifer Kreft}
\author{Matthias Schr\"oter}
\author{J. B. Swift}
\author{Harry L. Swinney}
\affiliation{Center for Nonlinear Dynamics and Department of Physics, University of Texas, Austin, Texas, 78712}

\date{\today}
\begin{abstract}
Molecular dynamics simulations are used to study fluidization of a
vertically vibrated, three-dimensional shallow granular layer. As
the container acceleration is increased above $g$, the granular
temperature and root mean square particle displacement increase,
gradually fluidizing the layer. For nearly elastic particles, or low
shaking frequencies, or small layer depths, the end of the
fluidization process is marked by an abrupt increase in the granular
temperature and rms particle displacement. The layer is then fully
fluidized since macroscopic, fluid-like phenomena such as convection
rolls and surface waves are observed. Increasing the total
dissipation (by either decreasing the restitution coefficient or
increasing the total number of particles) decreases the increase in
granular temperature and rms particle displacement at fluidization,
and shifts the increase to higher accelerations. Increasing the
frequency also decreases the magnitude of the jump, and shifts the
change to lower accelerations.

\end{abstract}

\pacs{45.70.-n,64.70.Dv,05.70.Fh, 07.05.Tp}

\maketitle

\section{Note from authors}
 After further investigations, we find that the results for the temperature and rms displacement at low $\Gamma$ depend on the functional form of the velocity dependence of the restitution coefficient.  Changing the dependence from $e=\mathrm{max} [e_0,1-(1-e_0)(v_{n}/\sqrt{gd})^{3/4}]$ to $e=\mathrm{max}[e_0,1-(1-e_0)(v_{n}/\sqrt{gd})^{1/5}]$ significantly reduces both $T_{kin}$ and $\Delta r_{rms}$ at low $\Gamma$.  The event-driven algorithm used in the present work has been tested and found to work well for situations where most collisions occur at high
velocities, but this algorithm is less accurate for situations where many collisions occur at low velocities, as in the present work. Hence the results we have presented should be checked with an model that is more accurate for low velocity collisions.

\section{Introduction}
Granular materials are collections of many dissipative particles,
which can behave as either a solid or a fluid \cite{jaeger:96}. To
maintain a fluid-like state, energy must be continually supplied to
the grains since it is dissipated during collisions. Often this
energy input is realized by shaking a container vertically; the
non-dimensional shaking acceleration is given by
$\Gamma=4\pi^2Af^2/g$, where $g$ is the acceleration due to gravity,
$A$ the shaking amplitude, and $f$ the shaking frequency. With
increasing $\Gamma$ the behavior of the granular layer changes from
solid-like to fluid-like. This fluidization does not occur abruptly
at a particular $\Gamma$ but rather is a process that develops over
a range in $\Gamma$. Even for $\Gamma < 1$, there is relative motion
of the particles \cite{poeschel:00,renard:01}, and the contact
forces between particles show rich dynamics \cite{hecke:05}. As
$\Gamma$ is increased above unity, the particles start to move
randomly on length scales small compared to their diameter. This has
been measured with laser speckle methods, where it was also found
that in deep layers, upper layers fluidize first and then the
fluidization proceeds downward \cite{kim:02} with increasing
$\Gamma$. The onset of macroscopically visible motion occurs at
$\Gamma \approx 2$, depending on the experimental conditions
\cite{melotrans}.

Quasi two-dimensional geometries allow the study of the fluidization
of  vertically vibrated layers using direct imaging. Studies of
horizontal monolayers have examined crystallization \cite{shattuck}
and the coexistence of different phases that occur during
fluidization \cite{olafsen1998, olafsen:2005, nie2000, prevost:2004,
sublimation}. In a vertical, initially crystalline layer a distinct
jump in the height of the center of mass of the particles was found
as the layer  fluidized \cite{bmelt}.

We present here results from fully three-dimensional molecular
dynamics simulations of a shallow granular layer. The process of
fluidization is accompanied by a smooth increase in temperature and
particle mean square displacement that culminates in an abrupt
increase for high coefficient of restitution, or low shaking
frequency, or small layer depth. We identify this abrupt change with
the end of the fluidization process.

\section{Molecular Dynamics Simulation}
The event driven molecular dynamics simulation described in
\cite{bizon} was used for this study. The number of frictional,
rotational spheres of diameter $d$ used was varied to determine the
dependence of fluidization on the layer depth, $N$. The total number
of particles was $N \times 248$, corresponding to a volume fraction
of 0.58 at rest. The container was a box $15d \times 15d \times
300d$, in the $x$,$y$, and $z$ directions, respectively, with the
$z$ direction being opposite to gravity. The bottom plate oscillated
sinusoidally, $z_P=Asin(2\pi ft)$. The non-dimensional frequency,
$f^*=f \sqrt{Nd/g}$, was varied from 0.08 to 0.25. Unless otherwise
noted, periodic boundary conditions were imposed in the $x$ and $y$
directions to avoid convection induced by frictional sidewalls
\cite{knight:93,knight:96,biot,wildman:01}. Surface waves were
prevented from forming by choosing the horizontal dimensions ($15d$)
to be smaller than the pattern wavelength, which was found to be
$39d$ in simulations in a large box for a layer with depth $N=3.6$,
frequency $f^*=0.15$, and $\Gamma = 2.3$.

The parameters characterizing the particles are the coefficient of
normal restitution $e$ (the ratio of the relative normal velocity
after collision to that velocity before collision), the coefficient
of friction $\mu$, and the tangential restitution $\beta$, which
gives the change in the relative surface velocity \cite{bizon,
walton}. For this study, $\mu$ and $\beta$ were fixed at $\mu=0.5$
and $\beta=0.35$, which were the values used to reproduce surface
wave patterns observed experimentally in oscillated layers of lead
spheres \cite{bizon}. The normal restitution $e$ depended upon the
relative normal velocity, $v_n$, of the colliding particles
according to:
$e=\mathrm{max}[e_0,1-(1-e_0)(v_{n}/\sqrt{gd})^{3/4}]$, as in
\cite{bizon}. The minimum value of $e$, $e_0$, was varied from
$0.98$ to $0.65$.

Simulations were started from random initial conditions at
$\Gamma=3$. After 300 plate oscillation cycles, $\Gamma$ was reduced
by 0.02. Particle positions and velocities were recorded in at the
phase where the plate was at its mean position and moving upwards.
This procedure was repeated with $\Gamma$ decreasing by 0.02 each
time until the particle collisions became too frequent for the
event-driven algorithm to be usable \cite{luding}. This limited our
investigations to $\Gamma \gtrsim$
1.6, but we found that the accessible range in $\Gamma$ included the
transition to full fluidization.

\section{Results}
\subsection{Fluidization}

\begin{figure} [hb]\centering

\includegraphics[width=8cm]{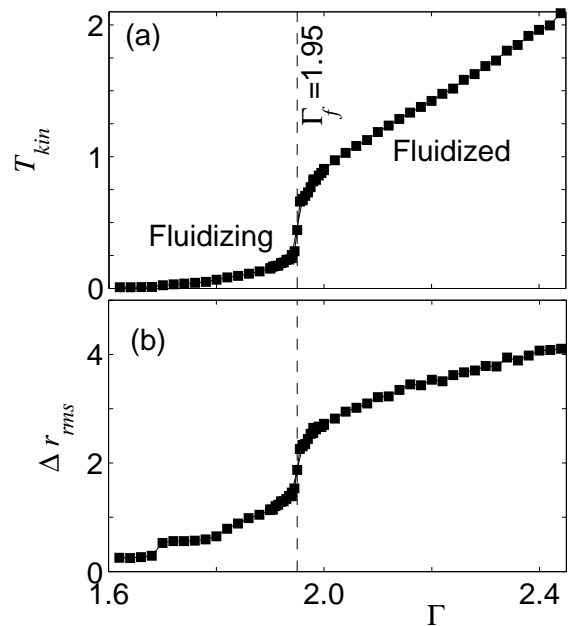} \caption{Comparison of
  different possible order parameters: (a) horizontal granular temperature, $T_{kin}$, and
(b) horizontal root mean square particle displacement in one cycle,
$\Delta r_{rms}$, as functions of $\Gamma$. Simulation parameters
were $e_0=0.98$, $N=3.6$, and $f^*=0.15$.} \label{fig1b}
\end{figure}

\begin{figure} [hb]\centering

\includegraphics[width=9cm]{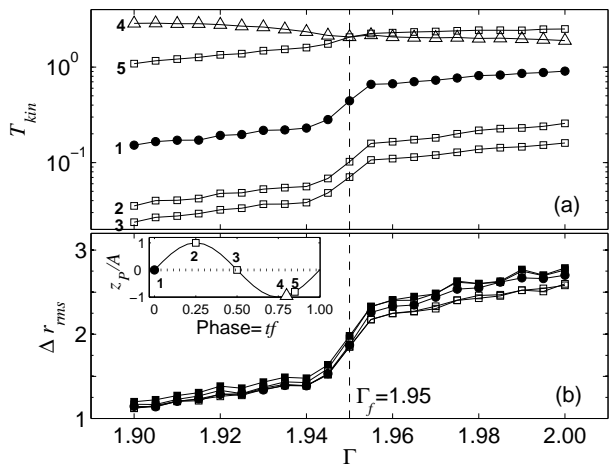} \caption{Phase dependence of the (a) granular temperature
and (b) root mean square particle displacement in one cycle as a
function of $\Gamma$. The inset in (b) shows the height of the plate
(normalized by the oscillation amplitude), $z_P/A$, as a function of
the phase $tf$  of the driving cycle. The temperature curves for
different phases in the cycle are similar except for phase four
(open triangles), which is influenced by the shock wave that passes
through the layer after the particles hit the plate \cite{Bougie:2002}. Filled circles indicate the phase used for the data
presented in other figures. Simulation parameters were $e_0=0.98$,
$N=3.6$, and $f^*=0.15$.} \label{fig2}
\end{figure}
\begin{figure} [hb]\centering

\includegraphics[width=9cm]{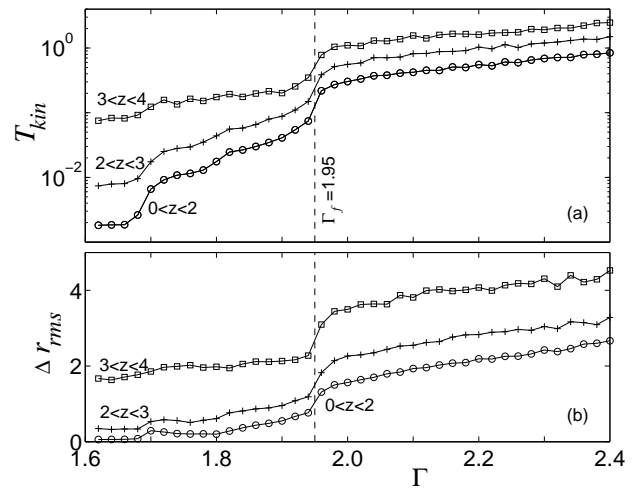}
\caption{Height dependence of (a) $T_{kin}$ and (b) $\Delta r_{rms}$
as functions of $\Gamma$, measured above the surface of the plate.
Fluidization completes at the same value of $\Gamma$ for each
height, but particles higher in the container are more mobile at
lower $\Gamma$. Here $e_0=0.98$, $N=3.6$, and $f^*=0.15$.}
\label{fig6} 
\end{figure}

The horizontal granular temperature $T_{kin}$ is the kinetic energy of the
relative random motion of the particles,
\begin{equation}
 T_{kin}=\frac{\left<\left(v_{x}-\left<v_{x}\right>\right)^2\right> + \left<\left(v_{y}-\left<v_{y}\right>\right)^2\right>}{2gd},
\end{equation}
where the averages are taken over all the particles and $T_{kin}$
has been made nondimensional by dividing it by the gravitational
potential energy of a single particle raised by a height equal to
its diameter. The averages are taken only in a horizontal plane to
minimize the influence of the vertical shock wave \cite{Bougie:2002}.
An example of the dependence of $T_{kin}$ on $\Gamma$ is shown in
Fig.\ref{fig1b}(a). A sharp increase in $T_{kin}$ is evident in the
range $1.94 < \Gamma < 1.955$.

We compare $T_{kin}$ with the root mean square displacement,
\begin{equation}
\Delta r_{rms}(t)=\frac{\sqrt{\left<\left(x(t)-x(0)\right)^2+\left(y(t)-y(0)\right)^2\right>}}{d}
\end{equation}
 where  $x(t)$ and $y(t)$ are the $x$ and $y$ positions of a particle at time $t$.
To minimize the impact of particles crossing the periodic boundary
on the measurement, the average is computed over all particles at
least $5d$ from the boundary at $t=0$. The rms displacement for
$\Delta t=1$ cycle(Fig.~\ref{fig1b}(b)) exhibits an increase at
$\Gamma_f$, just as found for $T_{kin}$ (Fig.\ref{fig1b}(a)).

We define as fluidization the process that begins when $\Gamma <1$
and ends at $\Gamma_f$, the acceleration where the abrupt increase
in $T_{kin}$ and $\Delta r_{rms}$ occurs. After the increase, the
layer is defined as vibrofluidized. This transition is history
independent and non-hysteretic: it occurs at the same $\Gamma$
whether decreasing $\Gamma$ from 3.00 or increasing from 1.84, for
step sizes that were as small as 0.01.

The granular temperature and density of vertically oscillated layers
depend on the phase of the plate oscillation because, after each
impact of the layer with the plate, a shock propagates upward
through the layer \cite{Bougie:2002}. The phase dependence of
$T_{kin}$ and $\Delta r_{rms}$ was examined for twenty equally
spaced phases throughout a cycle, and all but one phase were found
to yield similar behavior (Fig.~\ref{fig2}). The one phase that
yielded a different dependence on $\Gamma$ is labeled 4 in
Fig.~\ref{fig2}; at this point in the cycle the plate is near its
minimum height and moving upward, which is where the shock has the
greatest effect~\cite{Bougie:2002}. However, the shock quickly
travels through the layer, and its effects are negligible for most
of the cycle. We therefore recorded data at the phase when the plate
was at its mean position and moving upward.

Just as the vertical forcing may add a dependence on the phase of
the driving cycle, it also creates gradients in the $z$ direction.
The top layers are more fluidized than the lower layers, as Fig.~
\ref{fig6} illustrates.  While the fluidization process proceeds
more rapidly for the topmost layers, it does not complete until all
layers have fluidized (\ref{fig6}). Similar behavior was found for
$f^*=0.25$.

\begin{figure} [hb]\centering

\includegraphics[width=9cm]{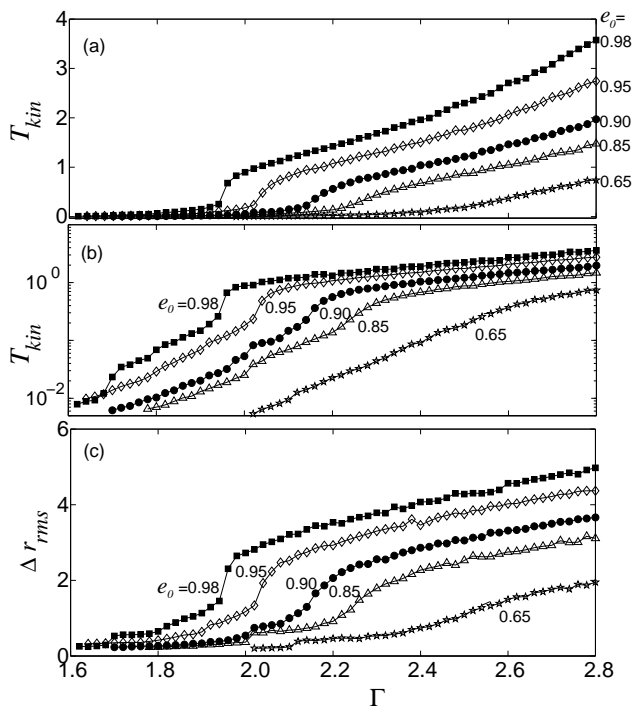} \caption{(a) Linear and
(b) logarithmic plots of the granular temperature, and (c) root mean
square displacement as a function of $\Gamma$ for different values
of the normal restitution coefficient.  The fluidization of the
layer is more abrupt for particles that are nearly elastic.
Simulation parameters were $f^*=0.15$ and $N=3.6$.} \label{fig1}
\end{figure}

\subsection{Dependence on System Parameters}
Changing the normal restitution coefficient has a dramatic effect on
the granular temperature and rms displacement, as shown in Fig.
~\ref{fig1}. The transition to the fully fluidized state is most
pronounced for our highest restitution value, $e_0=0.98$, and
becomes less obvious as $e_0$ is decreased; for the lowest
restitution value examined, $e_0=0.65$, no transition is
discernible. As $e_0$ is decreased, the fully fluidized state is
reached at higher $\Gamma$: for $e_0=0.98$, $\Gamma_f=1.95$; for
$e_0=0.85$, $\Gamma_f \approx 2.3$.

\begin{figure} [hb]\centering

\includegraphics[width=9cm]{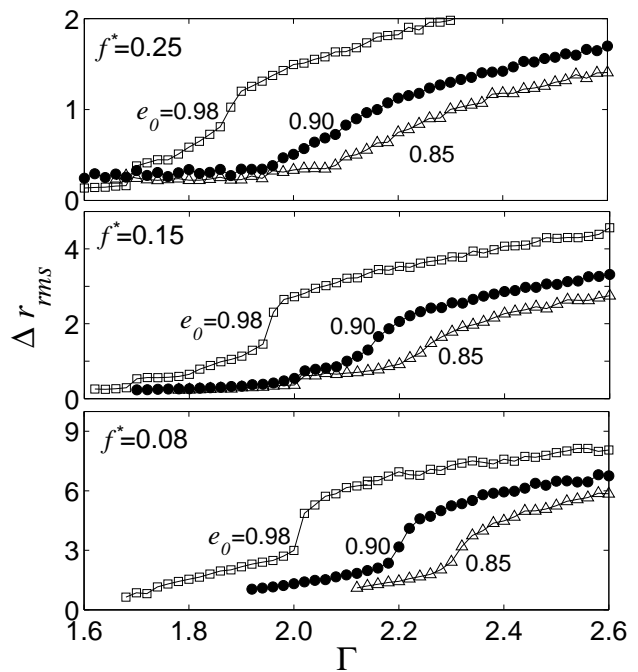}
\caption{The effect of varying the frequency on the $\Gamma$
dependence of $\Delta r_{rms}$. The change signaling the end of the
fluidization process becomes more pronounced as the frequency is
decreased. Here $N=3.6$.} \label{fig3}
\end{figure}

\begin{figure} [hb]\centering

\includegraphics[width=9cm]{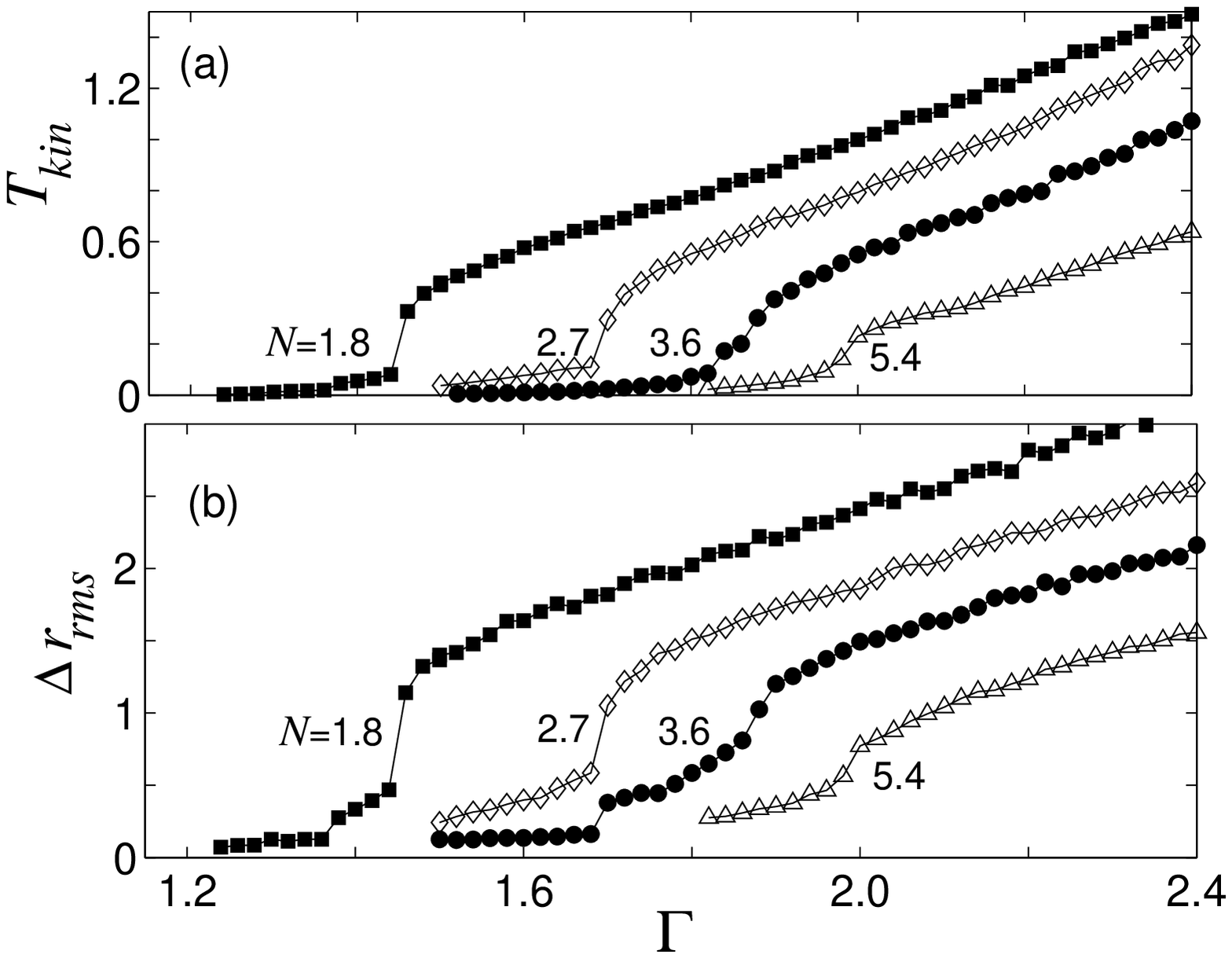} \caption{The effect
of varying the layer depth $N$ on the $\Gamma$ dependence of
$T_{kin}$ and $\Delta r_{rms}$. As the layer becomes thicker, the
change signaling the end of the fluidization process becomes less
pronounced. Simulation parameters were $f^*=0.25$ and $e_0=0.98$. }
\label{fig4}
\end{figure}

The dependence of $\Delta r_{rms}$ on $\Gamma$ and $e_0$ is compared
for three frequencies in Fig.~\ref{fig3}. Increasing the frequency
makes the fluidization process more gradual and decreases
$\Gamma_f$. For example, for $f^*=0.08$ (with $e_0=0.98$),
$\Gamma_f=2.0$, while if $f^*$ is increased to 0.25,
$\Gamma_f=1.85$.

Increasing the layer depth leads to a less pronounced increase in
$T_{kin}$ and $\Delta r_{rms}$, and the increase is shifted to
higher $\Gamma$ (Fig.~\ref{fig4}). For $N=1.8$, the end of the
fluidization process occurs at $\Gamma_f=1.48$, while for $N=5.4$
the change is smaller and $\Gamma_f \approx 2$. Thus, as the total
dissipation of the system increases, either by decreasing the
restitution coefficient or by increasing the number of collisions
(by adding more particles), the change from the fluidizing to the
fluidized state becomes less pronounced and $\Gamma_f$ increases.

After the layer has completed the fluidization process, fluid-like
phenomena appear in addition to the increases in $T_{kin}$ and
$\Delta r_{rms}$. For example, as is well known, surface waves
appear at $\Gamma \approx 2.5$,
\cite{bizon,umbanhowar:00,Melo:1994}, where $ \Gamma_f \approx 2$
\cite{melotrans}. We have done some simulations in a container large
enough to accommodate waves and have found that waves emerge in a
thin layer at about $\Gamma_f$, while for a deeper layer, pattern
onset occurs for $\Gamma > \Gamma_f$ (see Table \ref{table1}).
Additional simulations were made for a container with solid
frictional lateral walls, and a single convection roll was found to
develop at $\Gamma=2.24$, the same $\Gamma$ at which the
fluidization process completes (Table \ref{table1}).

\begin{table}
\caption{$\Gamma_{onset}$, the onset of fluid-like phenomena, is
found to be greater than or equal to $\Gamma_{f}$, the beginning of
the fully fluidized state.} \label{table1}
\begin{tabular}{|c|c|c|c|c|c|c|}
\hline
Phenomenom&$\Gamma_{onset}$&$\Gamma_{f}$&$x \times y$&$e_0$&$N$&$f^*$\\
\hline
\hline
Waves&$\approx 2.3$&1.95&$30 \times 30$&0.98&5.4&0.22\\
\hline
Waves&$\approx 2.3$&$\approx 2.3$&$30 \times 30$&0.85&3.6&0.15\\
\hline
\hline
Convection& 2.24 & 2.24&$15 \times 15$&0.85&3.6&0.15\\
\hline

\end{tabular}
\end{table}
\subsection{Comparison with Experiment}
Comparing these simulations to experiment is problematic.
Measurements of the granular temperature or rms displacement in a
three-dimensional experiment are challenging since the trajectories
of the grains in the bulk of the layer can not be recorded with just
a simple video camera. Recently speckle visibility has been
developed to measure the granular temperature in a 3D sample
\cite{dixon:03} and should be utilized to fully explore this
process.

Umbanhowar and Swinney investigated the
transition to surface waves for systems with similar layer depths and frequency using the pressure exerted by the layer on the bottom plate.
They therefore recorded data at a different point in the driving cycle than was used in this work.
In addition, the fluidization process may have completed at a lower
$\Gamma$ value than they report for the pattern transition as in our
own work (Table ~\ref{table1}) \cite{umbanhowar:00}.

Two groups have studied fluidization in deep layers ($N=15d$ in
\cite{melotrans} and $N>20d$ in \cite{kim:02}) and have reported
that the fluidization of the layer begins at the top and proceeds
downward as $\Gamma$ is increased. Evidence of this top down
fluidization can be seen in Fig. \ref{fig6}. Particles higher in the
container are in a more fluidized, but fluidization is completed at
the same value of $\Gamma$ for particles at all heights studied.

\section{Discussion}
We have shown that in shallow layers of nearly elastic particles,
abrupt increases  of both the granular temperature and the root mean
square displacement indicate that the process of fluidization is
completed. If the dissipation in the system is increased by adding
more particles or decreasing the coefficient of restitution, the
increases in granular temperature and rms displacement at full
fluidization become more and more gradual.

While the granular temperature and root mean square displacement are
good indicators of the onset of the vibrofluidized state, the
question remains if there exists a single control parameter for the
fluidization of a vertically oscillated granular sample. The
non-dimensional acceleration $\Gamma$ is not satisfactory since
layers with different characteristics fluidize at different values
of $\Gamma$. The value of the temperature $T_{kin}$ depends on
the phase of the cycle, and both the rms
displacement and $T_{kin}$ at the onset of the vibrofluidized state depend
on granular parameters such as coefficient of restitution and layer depth. Thus they do not make good measures.
Others have suggested $v_0=A(2\pi f)$, but this measure does not
collapse our data since we find a different dependence on $f^*$ than
that reported in \cite{bmelt}. We hope that our results will inspire
further work on the process  of fluidization in three-dimensional
granular layers.

The authors thank William McCormick and Chris Kr\"ulle for helpful
discussions. This work is supported by the Robert A. Welch
Foundation grant F-0805.

\bibliographystyle{apsrev}
\bibliography{thesis1,bne,bizon,bougie,cnldpubs,goldman,gran_temp,granular,kreft,lattice,melo,moon,onset,sublimation,thesis2}

\end{document}